\documentclass[11pt]{article}
\usepackage{amssymb}
\usepackage{amsmath}
\usepackage[all]{xy}
\usepackage{graphicx}
\usepackage{braket}
\usepackage{xcolor}

\title{Quantum Probability:\\a reliable tool for an agent or\\a reliable source of reality?}

\author{{\sc C. de Ronde}$^{1,2}$ {\sc H. Freytes}$^{3}$ and {\sc G. Sergioli}$^{3}$}
\date{}

\usepackage[margin=2.3cm]{geometry}

\begin{document}

\bibliographystyle{plain}
\maketitle

\begin{center}
\begin{small}
1. Philosophy Institute Dr. A. Korn (UBA-CONICET)\\
2. Center Leo Apostel for Interdisciplinary Studies\\Foundations of the Exact Sciences (Vrije Universiteit Brussel)\\
3. University of Cagliari
\end{small}
\end{center}

\bigskip

\begin{abstract}
\noindent In this paper we attempt to analyze the concept of quantum probability within quantum computation and quantum computational logic. While the subjectivist interpretation of quantum probability explains it as a reliable predictive tool for an agent in order to compute measurement outcomes, the objectivist interpretation understands quantum probability as providing reliable information of a real state of affairs. After discussing these different viewpoints we propose a particular objectivist interpretation grounded on the idea that the Born rule provides information about an intensive realm of reality. We then turn our attention to the way in which the subjectivist interpretation of probability is presently applied within both quantum computation and quantum computational logic. Taking as a standpoint our proposed intensive account of quantum probability we discuss the possibilities and advantages it might open for the modeling and development of both quantum computation and quantum computational logic. 
\end{abstract}
\begin{small}

\end{small}

\newtheorem{theo}{Theorem}[section]
\newtheorem{definition}[theo]{Definition}
\newtheorem{lem}[theo]{Lemma}
\newtheorem{met}[theo]{Method}
\newtheorem{prop}[theo]{Proposition}
\newtheorem{coro}[theo]{Corollary}
\newtheorem{exam}[theo]{Example}
\newtheorem{rema}[theo]{Remark}{\hspace*{4mm}}
\newtheorem{example}[theo]{Example}
\newcommand{\proof}{\noindent {\em Proof:\/}{\hspace*{4mm}}}
\newcommand{\qed}{\hfill$\Box$}
\newcommand{\ninv}{\mathord{\sim}} 
\newtheorem{postulate}[theo]{Postulate}

\bigskip

\bigskip

\section*{Introduction}

The notion of probability has been, since its origin, at the center of foundational debates in  Quantum Mechanics (QM). As it is well known, the so called ``ignorance interpretation'' applied in classical Kolmogorovian probability theory was found out to be incompatible with the formalism of the theory of quanta. The impossibility to interpret quantum probability in classical terms, as providing partial information of an actually existent, but unknown, state of affairs played an important role within many philosophical discussions and debates taking place between the founding fathers of the theory during the first half of the last century. Today, the problem remains still open and there is no general consensus inside the philosophical community regarding the meaning of the famous (probabilistic) Born rule. In fact, rather than getting closer to a consensus, the number of interpretations of probability has multiplied through the course of the years. In this article we attempt to analyze probability by discussing two general viewpoints. While the first considers quantum probability as a reliable predictive tool used by agents in order to compute measurement outcomes, the second understands it as providing a reliable informational source of a real state of affairs ---as theoretically described by QM. We will investigate how these different accounts of the Born rule provide a radically different perspective regarding the modeling of quantum computational processes.

The paper is organized as follows. In section 1, we provide a short introduction to quantum probability and its orthodox account. Section 2 analyses the subjectivist interpretation of quantum probability which argues that the Born rule can be understood as a reliable tool that can be used by an agent in order to bet on possible measurement outcomes. Section 3 introduces a new objectivist interpretation of quantum probability as providing a reliable informational source of a real intensive state of affairs theoretically described by QM. Section 4 provides a general analysis of the different ways in which the notion of probability is used in quantum computation and quantum computational logic. Taking as a standpoint our objectivist interpretation of probability, in section 5, we present a new intensive approach to the modeling of quantum computational processes which provides several advantages with respect to the orthodox modeling. Finally, we present the conclusions of the paper.

\section{Quantum probability}

Since Born's 1926 interpretation of the quantum wave function, $\Psi$, probability has become one of the key notions in the description of quantum phenomena. The rule provided a reliable account of statistical outcomes; however, contrary to the expectations of many, it did not explain how single results suddenly appeared. In his paper, Born formulated the now-standard interpretation of $\psi (x)$ as encoding a probability density function for a certain particle to be found at a given region. For a state $\psi$, the associated probability function is $\psi^{*} \psi$, which is equal to $|\psi (x)|^2$. If $|\psi (x)|^2$ has a finite integral over the whole of three-dimensional space, then it is possible to choose a normalizing constant. The probability that a particle is within a particular region $V$ is the integral over $V$ of $|\psi (x)|^2$. However, even though this interpretation worked fairly well, it soon became evident that the concept of probability in the new theory departed from the physical notion considered in classical statistical mechanics as {\it lack of knowledge} about a preexistent (actual) state of affairs described in terms of definite valued properties. According to Born himself:
\begin{quote}
\noindent {\small ``Schr\"{o}dinger's quantum mechanics [therefore] gives quite a definite answer to the question of the effect of the collision; but there is no question of any causal description. One gets no answer to the question, `what is the state after the collision' but only to the question, `how probable is a specified outcome of the collision'. Here the whole problem of determinism comes up. From the standpoint of our quantum mechanics there is no quantity which in any individual case causally fixes the consequence of the collision; but also experimentally we have so far no reason to believe that there are some inner properties of the atom which condition a definite outcome for the collision. [...] I myself am inclined to give up determinism in the world of the atoms. But that is a philosophical question for which physical arguments alone are not decisive.'' \cite[p. 57]{WZ}}
\end{quote}

Albert Einstein was clearly mortified by the lack of intuitive ({\it anschaulich}) spacio-temporal content of the theory of quanta and the introduction of the strange role played by measurement. Indeed, according to the projection postulate, explicitly included by von Neumann in his famous axiomatic formulation of QM, a sudden ``collapse'' of the quantum wave function is induced every time a subject performs a measurement. Einstein would argue ---almost in complete solitude--- that ``the moon has a definite position whether or not we look at the moon. [...] Observation cannot {\it create} an element of reality like position, there must be something contained in the complete description of physical reality which corresponds to the {\it possibility} of observing a position, already before the observation has been actually made.'' His unease with QM was mainly related to the consideration of a theory as an algorithmic recipe: ``I still believe in the possibility of a model of reality [...] a theory which represents things themselves and not merely the probability of their occurrence'' \cite{Einstein34}. At the very opposite side, the centrality of the subject within the new theory of quanta was completely embraced by Niels Bohr who even argued that the most important (epistemological) lesson of QM is that, we subjects, are not merely spectators  but also actors in the great drama of (quantum) existence. 

\smallskip

In the history of physics the development of probability took place through a concrete physical problem considered in the 18th Century. The physical problem was how to characterize a state of affairs even though the possessed knowledge of it was incomplete. Or in other words, how to deal with gambling. This physical problem was connected later on to a mathematical theory developed by Laplace and others. But it was only at the beginning of 20th Century that Kolmogorov was able to axiomatize this mathematical theory \cite{Kolmogorov}. Anyhow, even though there are still today many interpretational problems regarding the physical understanding of classical probability, when physicists talk about probability in statistical mechanics they discuss about the (average values of) properties of an uncertain ---but existent---  state of affairs.\footnote{In this respect it is important to remark that the orthodox interpretation of probability in terms of relative frequencies refers to `events' and not to `properties of a system'. This frequentist interpretation of probability is not necessarily linked to a realistic physical representation but rather supports an empiricist account of the observed measurement results.} This is why the problem to determine a definite state of affairs in QM ---the sets of definite valued properties which characterize the quantum system--- poses also problems to the interpretation of probability within the theory itself. As noticed by Schr\"odinger in a letter to Einstein:
\begin{quote}
\noindent {\small ``It seems to me that the concept of probability [related to quantum theory] is terribly mishandled these days. Probability surely has as its substance a statement as to whether something {\small {\it is}} or {\small {\it is not}} the case ---an uncertain statement, to be sure. But nevertheless it has meaning only if one is indeed convinced that the something in question quite definitely {\small {\it is}} or {\small {\it is not}} the case. A probabilistic assertion presupposes the full reality of its subject. \cite[p. 115]{Bub97}}
\end{quote}

\noindent Schr\"odinger \cite[p. 156]{Schr35} knew very well that in QM it is not possible to assign a definite value to all properties of a quantum state. As he remarked: ``[...] if I wish to ascribe to the [quantum mechanical] model at each moment a definite (merely not exactly known to me) state, or (which is the same) to all determining parts definite (merely not exactly known to me) numerical values, then there is no supposition as to these numerical values to be imagined that would not conflict with some portion of quantum theoretical assertions.'' This impossibility would be exposed three decades after in formal terms by Simon Kochen and Ernst Specker in their now famous theorem \cite{KS} ---to which we shall return in the next section. In 1981, Luigi Accardi proved that there is a direct link between Bell inequalities and probability models \cite{Accardi82}. The theorem of Accardi states that any theory which violates Boole-Bell inequalities\footnote{As remarked by Itamar Pitowsky \cite[p. 95]{Pitowsky94}: ``In the mid-nineteenth century George Boole formulated his `conditions of possible experience'. These are equations and inequalities that the relative frequencies of (logically connected) events must satisfy. Some of Boole's conditions have been rediscovered in more recent years by physicists, including Bell inequalities, Clauser Horne inequalities, and many others.''} has a non-Kolmogorovian probability model. Consequently, QM ---which is known to violate these inequalities--- possesses a probability model which cannot be related to our classical understanding of reality (see for a detailed analysis \cite{Svozil17}). Of course, a possibility ---that triggered Bell's investigation--- is to follow the path inaugurated by Bohmian theory and change the formalism of QM in order to retain the classical representation of physics in terms of particles with well defined positions in space-time. As argued by Bohm: 
\begin{quotation}
\noindent {\small ``[...] in the usual interpretation two completely
different kinds of statistics are needed. First, there is the
ordinary statistical mechanics, which treats of the distortion of
systems among the quantum states, resulting from various chaotic
factors such as collisions. The need of this type of statistics
could in principle be avoided by means of more accurate measurements
which would supply more detailed information about the quantum
state, but in systems of appreciable complexity, such measurements
would be impracticably difficult. Secondly, however, there is the
fundamental and irreducible probability distribution,
$P(x)=|\psi(x)|^{2}$ [...]. The need of this type of statistics
cannot even in principle be avoided by means of better measurements,
nor can it be explained in terms of the effects of random collision
processes. [...] On the other hand, the causal interpretation
requires only one kind of probability. For as we have seen, we can
deduce the probability distribution $P(x)=|\psi(x)|^{2}$ as a
consequence of the same random collision processes that give rise to
the statistical distributions among the quantum states." \cite[p. 456]{Bohm53}}
\end{quotation}

Changing the perspective of the problem but staying close to the orthodox mathematical formalism it is also possible to understand quantum probability in terms of the ignorance, not of a real existent state of affairs, but of an agent's knowledge of the future actualization of a measurement outcome. This account of probability, which goes under the name of the subjectivist interpretation, has provided a ground in the last decades for two of todays most popular interpretations of QM: Quantum Bayesianism and the many worlds interpretation.

\section{Quantum probability as an agent's reliable tool (in order to compute future measurement outcomes)} 

Many leading figures like Einstein, Schr\"odinger, Pauli and Heisenberg were certainly interested in metaphysics and the referential relation of physical theories to nature and reality. However, the pragmatic and instrumentalist understanding of science radicalized after the success of the Manhattan project in the IIWW closed the door to foundational questions. This was at least until in the 1980s Aspect and his group were able to prove the violation of the Boole-Bell inequality in an EPR type experiment. As remarked by Jeff Bub \cite{Bub17}: ``[...] it was not until the 1980s that physicists, computer scientists, and cryptographers began to regard the non-local correlations of entangled quantum states as a new kind of non-classical resource that could be exploited, rather than an embarrassment to be explained away.'' This new situation reopened the door to foundational problems which, in turn, placed the meaning of quantum probability ---once again--- as a central topic of debate. However, the analysis of quantum probability was still surrounded by the widespread anti-realist attitude towards the understanding of QM. As explained by Healey \cite{Healey}: ``[T]he instrumentalist takes theoretical statements to be neither true nor false of the world, science to aim only at theories that accommodate and predict our observations, and theories even in mature science to have given us increasingly reliable and accurate predictions only of things we can observe.'' Instrumentalism, as an extreme form of anti-realism, holds that QM is not a reliable theory in order to describe objective physical reality (objective reliability), it can be considered reliable as an abstract mathematical symbolism which only accounts for subjective observations (subjective reliability). In the year 2000, exactly one century after the beginning of the quantum voyage, in a paper entitled {\it Quantum Theory Needs no `Interpretation'} Christopher Fuchs and Asher Peres took the instrumentalist standpoint explicitly inside the philosophical debate about the meaning and reference of QM. According to them:  ``[...] quantum theory does not describe physical reality. What it does is provide an algorithm for computing probabilities for the macroscopic events (`detector clicks') that are the consequences of experimental interventions. This strict definition of the scope of quantum theory is the only interpretation ever needed, whether by experimenters or theorists.'' \cite[p. 70]{FuchsPeres00} The ideas presented in this paper were further developed by Fuchs and R\"udiger Schack taking as a basis the Bayesian interpretation of probability \cite{QBism13}. Due to their Bayesian standpoint in order to account for quantum theory, they called their approach quantum Bayesianism, or in short: QBism.

David Mermin, an old friend of Peres, after having tried with no success to develop a realist account of QM \cite{Mermin98} recently converted into QBism. Together with Fuchs and Shack \cite[ p. 750]{QBism13} they wrote: ``A measurement in QBism is more than a procedure in a laboratory. It is any action an agent takes to elicit a set of possible experiences. The measurement outcome is the particular experience of that agent elicited in this way. Given a measurement outcome, the quantum formalism guides the agent in updating her probabilities for subsequent measurements.'' Indeed, as QBists make explicitly clear: ``A measurement does not, as the term unfortunately suggests, reveal a pre-existing state of affairs.'' Measurements are personal, individual and QM is a reliable ``tool'' for the ``user''  ---as Mermin prefers to call the ``agent'' \cite{Mermin14}. Just like a mobile phone or a laptop, QM is a tool that we subjects use in order to organize our experience in a reliable manner.
\begin{quotation}
\noindent {\small ``QBist takes quantum mechanics to be a personal mode of thought ---a very powerful tool that any agent can use to organize her own experience. That each of us can use such a tool to organize our own experience with spectacular success is an extremely important objective fact about the world we live in. But quantum mechanics itself does not deal directly with the objective world; it deals with the experiences of that objective world that belong to whatever particular agent is making use of the quantum theory.'' \cite[p. 751]{QBism13}}
\end{quotation}

But the subjectivist interpretation of probability has not been used exclusively by anti-realist approaches to QM. It has been also used by the now popular Many Worlds realist interpretation. Attempting to make sense of the effectiveness of quantum computational processes David Deutsch proposed the existence of ``parallel'' worlds in which computations actually take place. By considering the theory of decision Deutsch proposed to explain the meaning of probability in purely subjective terms and even claimed to have derived Born's rule form subjective likehood \cite{Deutsch99}. This idea was developed by David Wallace who analyses the rational behavior of agents within the so called ``multiverse'' \cite{Wallace07}.\footnote{The subjectivist derivation of the Born rule by Deutsch and Wallace has received many criticism (e.g., \cite{DawidThebault14, Dieks07}).} Adrian Kent summarizes the subjectivist approach in the following terms: 
\begin{quotation}
\noindent {\small``One idea lately advocated by David Deutsch and David Wallace of the University of Oxford is to try to use decision theory, the area of mathematics that concerns rational decision-making, to explain how rational people should behave if they believe they are in a branching universe. Deutsch and Wallace start from a few purportedly simple and natural technical assumptions about the preferences one should have in a branching world and then claim to show that rational Everettians should behave as though they were in an uncertain probabilistic world following the statistical laws of quantum theory, even though they believe their true situation is very different.'' \cite[p. 6]{Kent14}}
 \end{quotation}

It is important to remark, that just like instrumentalists who detach the question of reference from the analysis of quantum probability, the working mathematician does not need to constrain himself to any set of metaphysical principles. Instead of wondering about the reference of a theory she is only worried about the internal structure and coherency of the mathematical theory she is dealing with. `Probability' is regarded by the mathematician as a `theory of mathematics' and in this sense departs from any conceptual physical understanding which relates the formal structure to the world around us. A mathematician thinks of a probability model as the set of axioms which fit a mathematical structure and wonders about the internal consistency rather than about how this structure relates and can be interpreted in relation to experience and physical reality. As noticed by Hans Primas:
\begin{quote}
\noindent {\small ``Mathematical probability theory is just a branch of pure
mathematics, based on some axioms devoid of any interpretation. In
this framework, the concepts `probability', `independence', etc. are
conceptually unexplained notions, they have a purely mathematical
meaning. While there is a widespread agreement concerning the
essential features of the calculus of probability, there are widely
diverging opinions what the referent of mathematical probability
theory is.'' \cite[p. 582]{Primas99}}
\end{quote}

\noindent The important point is that when a mathematician and a physicist talk about `probability' they need not refer to the same concept. While for the mathematician the question of the relation between the mathematical structure of probability and experience plays no significant role, for the physicist who assumes a realist stance the question of probability is {\it necessarily} linked to experience and the representation of physical reality.

The fact that the orthodox formalism of QM relates to a non-Kolmogorovian probability model is not such a big issue from an instrumentalist perspective nor from a purely mathematical viewpoint. For the instrumentalist, who is only worried about the predictive effectiveness of the mathematical model this point might be regarded as irrelevant. The same situation happens with many mathematicians working with these probability structures who do not get astonished in any way by them. However, from a viewpoint which understands ---following Einstein--- that a physical theory must be capable of providing a conceptual representation of physical reality, the question which arises is very deep, namely, what is the meaning of a concept of probability which does not talk about the degree of knowledge of an actual state of affairs? In this case, in what sense can probability be considered as being objectively reliable? From our perspective, if such a question is not properly acknowledged, the statement ``QM is a theory about probabilities'' looses all physical content. It might be regarded as either an obvious mathematical statement with no interest ---it only states the well known fact that in QM there is a (non-Kolmogrovian) probability measure assigned via Gleason's theorem\footnote{The Born rule can be considered as a particular application of the Gleason's Theorem. More precisely, Gleason's theorem proves that the Born rule follows naturally from the structure obtained by the lattice of events in a complex Hilbert space. However, Born's rule can be also derived without considering Gleason's theorem (see \cite{AuffevesGrangier15}).}--- or a meaningless physical statement, since we do not know what quantum probability signifies in terms of a physical concept. If we are to understand QM as a physical theory, and not merely as an algorithmic tool which predicts measurement outcomes, it is clear that we still need to supply a link between the mathematical formalism and a set of adequate physical concepts which are capable of providing a coherent account of quantum phenomena. As we might recall Heisenberg's \cite[p. 264]{Heis73} remark: ``[t]he history of physics is not only a sequence of experimental discoveries and observations, followed by their mathematical description; it is also a history of concepts.  For an understanding of the phenomena the first condition is the introduction of adequate concepts. Only with the help of correct concepts can we really know what has been observed.''

\section{Quantum probability as providing reliable objective information (of an intensive state of affairs)}

Albert Einstein was clearly uncomfortable with the idea of a purely probabilistic theory that would not be capable to explain, in a reliable manner, why an atom would suddenly choose to decay (or not). A probabilistic description seemed to clash with an objective account of a state of affairs. But is this really the case? Is there no way to consider QM as describing an objective state of affairs in probabilistic terms? In \cite{deRonde16a}, it was argued that one can do so, at the price of giving up the classical representation of reality in terms of actual (binary) properties. If we pay the price of abandoning the metaphysical picture provided by actualism and we shift to a description of physical reality in potential terms one can still interpret quantum probability as providing reliable objective information of a (potential) state of affairs without ---like in the case of Bhomian mechanics---  changing the mathematical formalism of the theory. Our approach also goes clearly against the orthodox positivist viewpoint which assumes that physical theories are only reliable mathematical devices capable of predicting the future observations of individual agents. Taking as a standpoint the orthodox mathematical formalism of QM, we have chosen to investigate the possibility of developing a truly non-classical representation which is able to explain in an {\it anschaulich} manner what QM is really talking about.\footnote{For a detailed discussion of the intuitive account of our approach we refer to \cite{deRondeMassri18a, deRondeMassri18b}.} So, instead of assuming a presupposed classical representation in terms of `systems' with definite `states' and `properties', we have argued that it is the Born rule which suggests the need to produce a deep reconsideration of the way in which physical reality can be theoretically represented. The key to the solution of this problem is to recognize that it is the probabilistic Born rule itself which supplies the invariant content of the theory.

\smallskip
\smallskip

\noindent {\it
{\bf Born Rule:} Given a vector $\psi$ in a Hilbert space, the quantity $\langle \psi| P | \psi \rangle$  allows us to predict the average value of (any) observable, represented by the hermitian operator $P$}.
 
\smallskip
\smallskip

\noindent Let us remark that this probabilistic statement is completely independent of the choice of any particular basis (or context); i.e., it is non-contextual. 

Orthodoxly, it is argued that one can describe `systems' with definite `states' and `properties'. This encapsulation of reality in terms of the classical paradigm, mainly due to Bohr's doctrine of classical concepts,\footnote{Bohr \cite[p. 7]{WZ} argued that ``it would be a misconception to believe that the difficulties of the atomic theory may be evaded by eventually replacing the concepts of classical physics by new conceptual forms.''} has blocked the possibility to advance in the development of a new conceptual scheme. This is what David Deutsch \cite{Deutsch04} has rightly characterized as ``bad philosophy''; i.e., ``[a] philosophy that is not merely false, but actively prevents the growth of other knowledge.'' Taking distance from the Bohrian prohibition to consider physical reality beyond the theories of Newton and Maxwell, and simply taking into account the mathematical invariance of the Born rule, it would be possible to extend the definition of {\it element of physical reality} proposed by Einstein, Podolsky and Rosen \cite{EPR} ---a notion which, as discussed in extreme detail by Franco Selleri and Gino Torizzi \cite{SelleriTorizzi81}, has played a significant role in the contemporary realist vs. anti-realist debate of QM. Taking into account the probabilistic nature of the Born rule, the original EPR definition can be naturally extended in the following manner (see for a detailed discussion \cite{deRonde16a}):

\smallskip
\smallskip

\noindent {\it {\bf Generalized element of physical reality:} If we can predict in any way (i.e., both probabilistically or with certainty) the value of a physical quantity, then there exists an element of reality corresponding to that quantity.}

\smallskip
\smallskip

\noindent Our definition implies a reconfiguration of the meaning of the quantum formalism and the type of predictions it provides. It also allows us to understand Born's probabilistic rule in a new light; not as providing reliable information about a (subjectively observed) measurement result, but rather, as providing reliable objective information about a theoretically described (potential) state of affairs. In this respect, objective probability does not mean that particles behave in an intrinsically random manner. Objective probability means that probability characterizes in a reliable manner an intrinsic feature of the objectively described (formally and conceptually) state of affairs. Thus, the reference of objective probability is completely independent of any subjective choice. This account of probability allows us to restore a representation in which there exists an objective state of affairs which is completely detached from the observer's choices to measure (or not) a particular property ---just like Einstein desired.\footnote{As recalled by Pauli \cite[p. 122]{Pauli94}: ``{\it Einstein}'s opposition to [quantum mechanics] is again reflected in his papers which he published, at first in collaboration with {\it Rosen} and {\it Podolsky}, and later alone, as a critique of the concept of reality in quantum mechanics. We often discussed these questions together, and I invariably profited very greatly even when I could not agree with {\it Einstein}'s view. `Physics is after all the description of reality' he said to me, continuing, with a sarcastic glance in my direction `or should I perhaps say physics is the description of what one merely imagines?' This question clearly shows {\it Einstein}'s concern that the objective character of physics might be lost through a theory of the type of quantum mechanics, in that as a consequence of a wider conception of the objectivity of an explanation of nature the difference between physical reality and dream or hallucination might become blurred.'' See also the discussion in \cite{SelleriTorizzi81}.} 

Our proposed interpretation of quantum probability considers the Born rule as providing always complete and certain knowledge of the theoretically described objective state of affairs; both in cases where the probability is equal to 1 or 0 and also in cases in which probability belongs to the interval $(0,1)$. As it is argued in \cite{deRondeMassri18a}, the shift from an actualist representation of physical reality to a potential representation implies the shift from a {\it binary valuation} of properties to an {\it intensive valuation} of projection operators. This also implies a shift from an actualist understanding of certainty, related exclusively to the values 0 and 1; to a generalized {\it intensive certainty}, which considers on equal footing all values of probability within the closed interval $[0,1]$. 
\begin{definition}
{\rm Let $\mathcal{H}$ be a Hilbert space and $\mathcal{G}(\mathcal{H})$  be the set of projectors. A {\it potential state of affairs}\footnote{A similar definition is discussed in \cite{Kalmbach}.} (or PSA for short) is a function $\Psi:\mathcal{G}(\mathcal{H})\to[0,1]$ from the graph of observables $\mathcal{G}(\mathcal{H})$
such that $\Psi(I)=1$ and 
\[
\Psi(\bigvee_{i=1}^{\infty} P_i)=
\sum_{i=1}^\infty \Psi(P_i)\]
for any piecewise orthogonal projections $\{P_i\}_{i=1}^{\infty}$.
The numbers $\Psi(P) \in [0,1]$, are called {\it intensities} or {\it potentia}
and the elements  $P \in \mathcal{G}(\mathcal{H})$ are called \emph{immanent powers}.
Hence, a PSA assigns in a non-contextual manner through the Born rule a potentia to each hermitian operator. From a mathematical viewpoint, each $P_i$ can be identified with the closed subspace given by the image of $P_i$. If one considers the closed subspace $\bf P$ generated by the family $\{Image\,\, P_i\}_{i\in\mathbb N}$, then $\bigvee_{i=1}^{\infty} P_i$ denotes the projector onto $\bf P$}.
\end{definition}

Let $\mathcal{H}$ be a separable Hilbert space, $\dim(\mathcal{H})>2$ and $\mathcal{G}(\mathcal{H})$  be the set of immanent powers with the commuting relation given by QM. Then, any positive semi-definite self-adjoint operator of the trace class $\rho$ determines in a bijective way a PSA $\Psi:\mathcal{G}\to [0,1]$. A PSA is a non-contextual map assigning to each hermitian operator $P_i$ a number $p_i$ computed via the Born rule; i.e., each $P\in \mathcal{G}(\mathcal{H})$ the value $\mbox{Tr}(\rho P)\in[0,1]$.\footnote{In order to prove that this assignment is bijective, let  $\Psi:\mathcal{G}(\mathcal{H}) \to[0,1]$ be a PSA. Then, by the Gleason's theorem, there exists a unique positive semi-definite self-adjoint operator  of the trace class $\rho$ such that $\Psi$ is given by the Born rule with respect to $\rho$. Hence our claim.} The problem of KS contextuality in relation to the definition of physical reality appears from the impossibility to have a {\it Global Binary Valuation} related to the elements that can be possibly measured. Given a $\Psi$ the KS theorem precludes the possibility to consider the state of affairs in terms of definite valued properties, 0 or 1, and thus an objective (non-relative) representation becomes impossible. However, by escaping the metaphysical constraints imposed by the actualist binary representation, our approach is able to provide a {\it Global Intensive Valuation} to all hermitian operators $P_i$ related to the different decompositions of $\Psi$, thus restoring the possibility of an objective (non-contextual) account of the orthodox formalism. Taking the Born rule as a standpoint, it is possible to consider a generalized notion of valuation which, going beyond the restrictive binary valuation imposed by actualist metaphysics, is able to escape the idea that quantum contextuality implies that observations necessarily change the state of the system. Instead of being considered in terms of the ontic incompatibility of properties, quantum contextuality becomes a natural expression of the epistemic incompatibility of measurements (see for a detailed discussion \cite[section 8]{deRondeMassri18a}).  

In this way, our approach can be derived in a very natural manner from the orthodox formalism and physical considerations. This also provides a guide regarding the mathematical elements that should be considered in the creation and development of an objective conceptual representation. There is of course a number of metaphysical questions that arise when considering a potential realm as a mode of existence. Some of them have been already addressed in \cite{deRondeMassri18a, deRondeMassri18b}. However, a more detailed analysis and comparison with some related approaches exceed the scope of the present paper which we leave for a future work.\footnote{It is important to notice that, since the notion of object is completely dropped, the intensive approach differs from those which consider {\it vague objects} such as Evans \cite{Evans78}. It also differs from those who understand potentiality as a teleological cause for an actuality to take place; e.g., Heisenberg's potentiality, Margenau's latencies and Popper's propensities. For a detailed discussion on the topic we refer the interested reader to \cite{deRonde17, deRonde19}.} In the present article we are interested in the  way in which probability is actually used within orthodox quantum computation and quantum computational logic and the possibilities that our intensive approach might open within these specific fields of research.

\section{The role of probability in quantum computation and quantum computational logic}

Quantum Computation (QC) and Quantum Computational Logic (QCL) are both grounded on the widespread orthodox understanding of QM in terms of `properties', `states' and `systems'. The idea is to look to classical mechanics and extend these notions in an operational manner. As we know, the notion of state of a physical system is familiar in classical mechanics. In this case, the initial conditions (the initial values of position and momentum) determine the solutions of the equation of motion of the system allowing to follow the system in different instants of time. For any value of time, the state is represented by a point in the phase space. But of course, within the quantum framework, the consideration of states requires a substantial modification. In the orthodox scheme the definition of quantum state depends explicitly on the concept of \emph{maximal quantum test}. Taking as a standpoint the idea that QM talks about properties and systems, we want to observe the properties of a quantum system that can possibly take $n$ different values. If the test you devise allows us to distinguish among $n$ possibilities, it is said that this is a \emph{maximal test}. A $n$-outcome measurement of those properties implements a maximal test. A test that gives only partial information is said to be a \emph{partial test}.\footnote{By \emph{partial test} we mean what Peres \cite[p. 29]{Peres} named \emph{incomplete test}, i.e. ``[...] is one where some outcomes are lumped together, for example, because the experimental equipment has insufficient resolution.''} If a quantum system is prepared in such way that one can devise a maximal test yielding with certainty a particular outcome, then it is said that the quantum system is in a \emph{pure state} (for a detailed description see \cite[p. 30, Postulate A]{Peres}). The pure state of a quantum system is described by a unit vector in a Hilbert space, $\Psi$, which in a specific basis ---in Dirac notation--- is denoted by $|\psi \rangle$. Now, if the maximal test for a pure state has $n$ possible outcomes, the state is described by a vector $\Psi$ in a $n$-dimensional Hilbert space. Any orthornormal basis represents a realisable maximal test. Suppose that we have a large number of similarly prepared systems, called an ensemble, and we test for the values of different measurable quantities like, e.g., spin, position, etc. In general, we postulate that, for an ensemble in an arbitrary state, it is always possible to devise a test that yields the $n$ outcomes corresponding to an orthonormal basis with definite probabilities. If the system is prepared in state $\ket{\psi}$, and a maximal test corresponding to a basis $\{\ket{e_1}, . . . , \ket{e_n}\}$ is performed, the probability that the outcome will correspond to $\ket{e_i}$ is given by $p_i(\ket{\psi})=|\bra {e_i}\ket{\psi}|^2$.

But in general, a quantum system is not in a {\it pure state}, it might be also a {\it mixed state} (formally defined below). According to the orthodox account, this may be caused, for example, by an inefficiency in the preparation procedure of the system, or because, in practice, systems cannot be completely isolated from the environment, undergoing decoherence of their states. It is important to remark that the distinction between pure and mixed states is grounded on the possibility of predicting with certainty a specific outcome of an observable. This, of course, goes back to EPR's notion of {\it element of physical reality} which also provides an operational definition in terms of the possibility of measuring with certainty a physical quantity \cite{EPR}.  As Aerts and Sassoli de Bianchi \cite{AertsSassoli17} argue: ``the notion of `element of reality' is exactly what was meant by Einstein, Podolsky and Rosen, in their famous 1935 article. An element of reality is a state of prediction: a property of an entity that we know is actual, in the sense that, should we decide to observe it (i.e., to test its actuality), the outcome of the observation would be certainly successful.''

The idea of quantum computation was introduced in 1982 by Richard Feynmann and remained primarily of theoretical interest until developments such as, e.g., Deutsch-Jozsa algorithm and Shor's factorization algorithm formulated in the mid-90s, triggered a vast domain of research. In a classical computer, the information is encoded in a series of bits that are manipulated by logical gates. After a suitable sequence of steps the output is produced. Standard QC is based on the idea that `quantum systems' described by finite dimensional Hilbert spaces ---specially $\mathbb C^2$--- can be used in analogous manner to classical bits. Analogously to the classical computation case, we consider quantum logical gates (hereafter {\it quantum gates} for short) acting on qubits. QC can simulate any computation performed by a classical system, but taking advantage of quantum superpositions and quantum contextuality they can speed up the processes and make algorithms more efficient.  

However, there are also interesting processes that cannot be represented as unitary evolutions. A typical example is what happens at the end of a computational process, when a non-unitary operation, a measurement, is produced, and the superposed state ``collapses'' to only one of the terms. This is understood orthodoxly as a path from a pure state to a mixture; i.e., a probability distribution over different pure states. Quantum mixtures are then interpreted, in analogous terms to classical mixtures, as providing the probabilities for such outcomes. Taking mixed states into account, several authors \cite{AKN, DGG, DF, FD, FSA, GUD1, TA} developed a generalized model of quantum computational processes where {\it pure states} are replaced by {\it mixed states}. This new model is known as quantum computation with mixed states. Let $\mathcal H$ be a complex Hilbert space. We denote by $\mathcal L(\mathcal H)$ the algebra of operators on $\mathcal H$. In the framework of quantum computation with mixed states, we regard a quantum state in a Hilbert space $\mathcal H$ as a density operator i.e., an Hermitian operator $\mathcal{L(H)}$ that is positive semidefinite and has unit trace. We indicate by $\mathcal D(\mathcal H)$ the set of all density operators in $\mathcal H$. A quantum operation is a linear operator from density operators to density operators such that $\forall \rho\in\mathcal D(\mathcal H) : \epsilon (\rho) =\sum_i A_i\rho A_i^{\dagger}$, where $A_i$ are operators satisfying $\sum_iA_i^{\dagger}A_i=I$ and $A_i^{\dagger}$ is the adjoint of $A_i$. In the representation of quantum computational processes based on mixed states, a quantum circuit is a circuit whose inputs and outputs are labeled by density operators, and whose gates are labeled by quantum operations. In terms of density operators, an $n$-qubit $\ket{\psi}\in\otimes^n\mathbb C^2$ can be represented as a matrix $\ket{\psi}\bra{\psi}$. Moreover, we can associate to any unitary operator $U$ on a Hilbert space $\otimes^m\mathbb C^2$ a quantum operation $\mathcal{O_U}$, such that, for each $\rho\in\mathcal D(\mathcal H)$, $\mathcal{O_U}(\rho)=\mathcal U\rho\mathcal U^{\dagger}$. Apparently, quantum computation with mixed states generalizes the standard model based on qubits and unitary transformations.

\subsection{Probability in quantum computation}

The standard orthonormal basis $\{\ket0, \ket 1\}$ of $\mathbb C^2$ (where $\ket0 = (1, 0)^{\dagger}$ and $\ket1 = (0, 1)^{\dagger}$) is called the logical (or computational) basis. Pure states $\ket{\psi}=c_0\ket{0}+c_1\ket{1}$ in $\mathbb C^2$ are called quantum bits (or \emph{qubits}, for short) and are coherent superpositions of the basis vectors with complex coefficients $c_0$ and $c_1$. The two basis-elements $\ket 0$ and $\ket 1$ are usually taken as encoding the classical bit values 0 and 1, respectively. By these means and accordingly with the Born rule, a probability value is assigned to a qubit as follows:
let us consider the qubit  $c_0\ket{0}+c_1\ket{1}$. Then its probability values are $p(\ket{0}) = |c_0|^2$ and $p(\ket{1}) = |c_1|^2$. Here, Born's rule is understood in purely epistemic terms as providing the probability to observe either the measurement outcome related to $\ket{0}$ or the measurement outcome related to $\ket{1}$. 

Both in classical and in quantum computation, a circuit is described in terms of a sequence of gates that transform an arbitrary state (input) into another state (output) \cite{KSV}. The Schr\"odinger equation describes the dynamic evolution of quantum systems, showing how the state $\ket{\psi_{t_0}}$ at the initial time $t_0$ evolves into another state $\ket{\psi_{t_1}}$ at the final time $t_1$ by the equation $\ket{\psi_{t_0}}\rightarrow \ket{\psi_{t_1}}=U\ket{\psi_{t_0}}$, where $U$ is a unitary operator that represents a reversible transformation. The Schr\"odinger equation is naturally applied in quantum information theory \cite{NC}. Indeed, quantum logical gates are unitary operators that describe the time evolution of a quantum input state. However, let us remark that to perform a quantum algorithm means to apply a sequence of quantum gates to a quantum input state and to make a measurement at the end of the process. In standard quantum computation, the evolution of a state during the computation (and before the measurement) is unitary; hence, the quantum gates that act before the measurement processes are unitary. But when we want to enclose in the mathematical framework also the measurement, then we must consider the extended model of quantum computation where quantum gates are described by quantum operations that are able to represent both the unitary processing of the quantum gates and the non-unitary process of measurement. 

The essential difference between classical and quantum information theory is given by the basic information quantity, that, in the classical framework, is stored by the {\it classical bit} while in quantum computation is given by the {\it quantum bit} introduced above \cite{DGS,NC}. 
The qubits allow to manipulate entangled states that play a crucial role for a large class of  computational problems whose computational complexity is exponential for a classical computer while is polynomial for a quantum one. It is commonly argued that qubits, because of the superposition principle, are able to store a very larger amount of information with respect to its classical counterpart. This is one of the main arguments used by researchers in the field in order to explain why quantum computation is more efficient than classical computation. However, quite recently, it has been also argued that ``contextuality supplies the magic for quantum computation'' \cite{Howard14}. Indeed, quantum contextuality must be also regarded as one of the main features present within QC. We will come back to these important aspects of QC in the following sections. 

The input of a quantum circuit is given by a composition of qubits that is mathematically represented by the tensor product operation. Hence, given $k$ qubits $\ket {x_1},\ket {x_2},\cdots,\ket {x_k}\in\mathbb C^2$ the input state given by an ensemble of $k$ qubits is given by $\ket {x_1}\otimes\ket {x_2}\otimes\cdots\otimes\ket {x_k}$ (that, for short, we call quantum register ---or \emph{quregister}--- and we indicate by $\ket {x_1 x_2 \cdots x_k}\in\otimes^k\mathbb C^2$). A quantum circuit is represented by the evolution of the input quregister under the application of some unitary quantum logical gate \cite{H, NC}. Also in a quantum computer it is possible to perform a {\it maximal test}: it consists in making a measurement on any output involved in the computation. By following the standard quantum scenario, by repeating several times the same computation it is possible to reach different results. For this reason the standard result of a computation process is given by running (and measuring) several times the same circuit. After repeating this procedure a sufficient member of times, what we get is a frequency distribution over a spectra of outcomes. And as we discussed above, this can be computed for each possible outcome in terms of the Born rule. Unlike the classical computation, in the quantum context a computation can provide different outputs for two different reasons. The first reason is given by the fact that the gates involved in the computation can generate a superposition. For instance, if we apply the Hadamard $H$ gate \cite{NC} to the input state $\ket 0$, in principle we have probability $\frac{1}{2}$ to get $\ket 0$ and $\frac{1}{2}$ to get $\ket 1$ as output (because $H\ket 0=\frac{1}{\sqrt 2}(\ket 0 + \ket 1)$). Hence, in principle, by applying $n$ times the computation of the Hadamard gate  on the input state $\ket 0$ we should obtain $\frac{n}{2}$ times the output $\ket 0$ and $\frac{n}{2}$ times the output $\ket 1$. Obviously, this behavior has no classical counterpart. While in CC at the end of a computation there is only one output determined by the algorithm, in QC even though there is only one outcome at the end of the computation the results can be as many as terms in the qubit and ---in general--- cannot be known before the measurement (or collapse) of the qubit.  

As an example, in Figure 1 we show a very trivial computation where the input is given by three qubits set in the state $\ket 0$ and the gates are: a double negation for the first qubit of the input, a double Hadamard gate for the second and an Identity gate for the third. All three qubits are measured at the end of the computation. By exploiting the \emph{Quantum Experience} resource \cite{De}, the computation process above has been made by a real IBM quantum computer and the same computational process is replicated by $1024$ different runs of the same computational configuration. We then obtain a distribution of frequency spanned on all possible outputs. Obviously, the most probable output (see Figure 2) is given by the state $\ket{000}$, but the other possible outcomes have also a non negligible probability. According to the orthodox viewpoint, this is given by some error that can occur during the computational process. Indeed, unlike the classical circuit, the real quantum circuit is particularly sensitive to the action of the environment (which can affect the computation); and this is the second reason that allows to have different outputs by running different times the same quantum circuit. Thus, the frequency distribution depicted in Figure 2 can be regarded as a kind of error; i.e., this frequency distribution can be seen as a tool that allows us to compute, if we apply the circuit of Figure 1, that given the stability of the circuit the state $\ket{00000}$ is the \emph{most probable} result if a measurement is performed.
\begin{figure}[htbp]\label{Circuit}
\centering
\includegraphics[scale=.20]{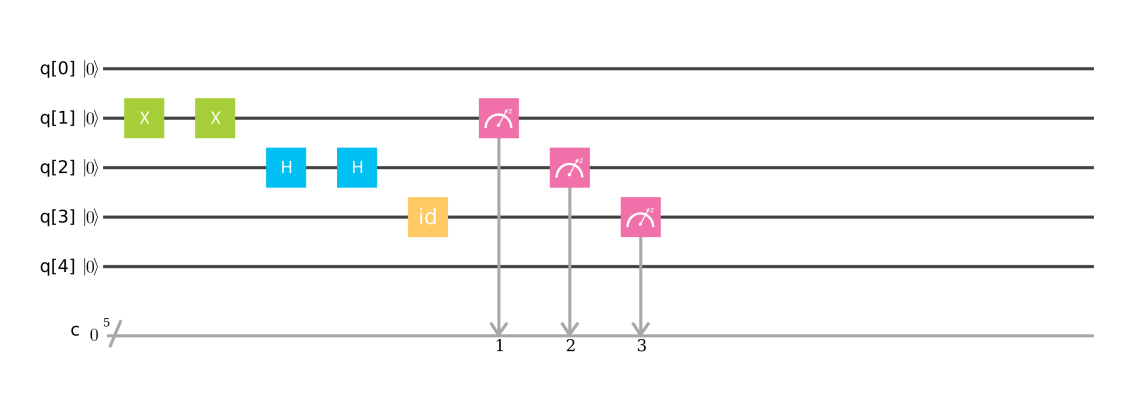}
\caption{Example of a quantum circuit}
\end{figure}
\begin{figure}[htbp]\label{Probability}
\centering
\includegraphics[scale=.28]{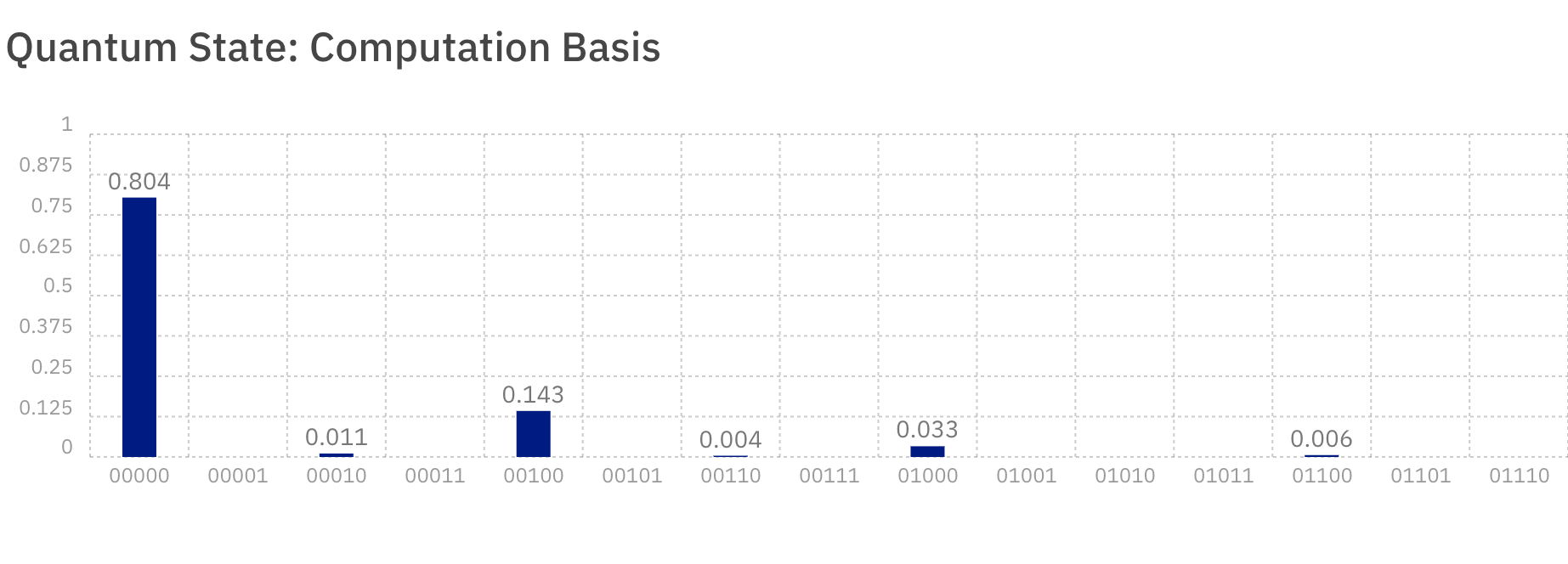}
\caption{Frequency distribution}
\end{figure}

To sum up, in a quantum circuit we have two types of statistical outcomes. One related to the fact a qubit might be in a superposition state and another statistical spectra related to the intrinsic stability or error present within the actual quantum circuit created in the lab. While in the first case the probability can be computed via the Born rule and is purely theoretical, in the latter case, just like in classical computations, the calculation is purely related to imperfections in the circuit which can be only known {\it a posteriori}, through the actual empirical test of the circuit.

\subsection{Probability in quantum computational logic}\label{QCL}

The theory of Quantum Computation has naturally inspired new forms of quantum logic, such as the so called \emph{Quantum Computational Logic} (QCL) \cite{DGG,SGP}. From a semantic point of view, any formula of the language in QCL denotes a piece of quantum information, i.e. a density operator living in a complex Hilbert space whose dimension depends on the linguistic complexity of the formula. Similarly, the logical connectives are interpreted as special examples of quantum gates. Accordingly, any formula of a quantum computational language can be regarded as a logical description of a quantum circuit. 

The kernel idea giving rise to QCL is the assignment of the truth value of a quantum state that represents a formula of the language. Conventionally, QCL assigns the truth value ``false" to the information stored by the qubit $\ket 0$ and the truth value ``true" to the qubit $\ket 1.$ Unlike classical logic, QCL turns out to be a \emph{probabilistic} logic, where the qubit $\ket\psi=c_0\ket 0 +c_1\ket 1$ logically represents a ``probabilistic superposition" of the two classical truth values, where, accordingly to the Born rule, the \emph{falsity} has probability $|c_0|^2$ and the \emph{truth} has the probability $|c_1|^2$. As in the qubit case, in the standard approach to QCL it is also defined a probability function $\texttt p$ that assings a probability value $\texttt p(\rho)$ to any density operator $\rho$ living in the space of the $n$-arbitrary dimensional density operators (we denote this space by $\mathcal D(\otimes^n\mathbb C^2)$). Intuitively, $\texttt p(\rho)$ is the probability that the quantum information stored by $\rho$ corresponds to a \emph{true} information that after a measurement we obtain either the outcome related to $\ket 0$ or $\ket 1$. 

In order to define the function $\texttt p$, we first need to identify in any space $\otimes^n\mathbb C^2$ the two operators $P_0^{(n)}=I^{(n-1)}\otimes(\ket0\bra0)$  and $P_1^{(n)}=I^{(n-1)}\otimes(\ket1\bra1)$ (with $I$ the identity operator) as the two special projectors that represent the \emph{falsity} and the \emph{truth} properties (in fact, outcomes), respectively. Before this, another step is crucial. In order to extend the definition of \emph{true} and \emph{false} from the space $\mathbb C^2$ of the qubits to the space $\otimes^n\mathbb C^2$ of the tensor product on $n$ qubits (i.e., on an arbitrary \emph{quregister}), the standard approach to QCL follows the convention that: a quregister $\ket x=\ket{x_1\dots x_n}$ is said to be \emph{false} if and only if $x_n=0$; conversely, it is said to be \emph{true} if and only if $x_n=1$. Hence, the truth value of a quregister only depends on its last component. On this basis, it is natural to define the property  \emph{falsity} (or \emph{truth}) on the space $\otimes^n\mathbb C^2$ as the projector $P_0^{(n)}$ (or $P_1^{(n)}$) onto the span of the set of all \emph{false} (or \emph{true}) registers. Now, accordingly with the Born rule, the probability that the state $\rho$ is \emph{true} is defined as: 
\begin{eqnarray}\texttt p(\rho)=Tr(P_1^{(n)}\rho).
\end{eqnarray}

In the language of QCL it is usual to distinguish between \emph{semiclassical} gates (called \emph{semiclassical} because, when they are applied to the elements of the computational basis $\mathbb B=\{\ket 0, \ket 1\}$, they replace the behavior of their corresponding classical logical gates) and \emph{genuinely} quantum gates (called \emph{genuinely} quantum because their application to the elements of the computational basis have no classical counterpart). The semiclassical gates usually involved in QCL are: the Identity $I$, the Negation $Not$, the control-negation (or Xor) $CNot$ and the Toffoli gate $T$, while the genuinely quantum gates are: the Hadamard gate $\sqrt I$ (also named square root of the identity) and the square root of the negation $\sqrt{Not}.$
In particular, the $T$ and the $Not$ gates allow us to provide a \emph{probabilistic} replacement of the classical logic in virtue of the following properties:
\begin{itemize}
\item $\texttt p(^\mathcal D{Not}(\rho))=1-\texttt p(\rho),\,  \text{for any} \rho\in\otimes^n\mathbb C^2;$
\item $\texttt p(AND(\rho,\sigma))=\texttt p(^\mathcal D{T}(\rho\otimes\sigma\otimes P_0))=\texttt p(\rho) \texttt p(\sigma), \, \text{for any} \rho,\sigma \in \otimes^n\mathbb C^2.$
\end{itemize}

\noindent Let us notice how the conjunction is obtained by the expedient to use the ternary Toffoli gate equipped by the projector $P_0$ that plays the role of an \emph{ancilla}. 

Taking this approach as a standpoint, and inspired by the intrinsic properties of quantum systems, the semantic of QCL turns out to be strongly non-compositional and context dependent \cite{DGLS}. This approach, that may appear \emph{prima facie} a little strange,   has the benefit of reflecting pretty well plenty of informal arguments that are currently used in our rational activity \cite{DGLNS}. A  detailed description of QCL and its algebraic properties is summarized in \cite{DGG,DGLS, DGS,LS}.

\section{An intensive probabilistic approach to QC and QCL}

The way in which probability is applied in QC and QCL rests on an operational definition of properties which is grounded in a direct manner on an operational understanding of {\it actual certainty} as the restriction to probability values equal to 1 (or 0). As we discussed above, this idea is captured explicitly within the operational definition of pure state and EPR's definition of {\it element of physical reality} \cite{EPR}. Departing itself from a reference to a really existent state of affairs, probability is understood as making reference to `the future certain prediction (probability = 1) of a measurement outcome'. In this way, both QC and QCL follow the subjectivist path of probability understood as a reliable tool for an agent (section 2). However, as we have discussed above (in section 3), there is another possible interpretation of quantum probability which, departing from the classical metaphysical representation in terms of `systems' and `properties', is still able to stays close to the orthodox formalism and provide a non-classical account of what QM is really talking about. This path allows to consider, through the reference to an {\it intensive} realm of existence, a notion of quantum probability which restores an objective reference to a state of affairs. There are several points which motivate such an intensive interpretation of probability to which we now turn our attention. 

First, the intensive interpretation of the Born rule allows to restore an objective, invariant and non-contextual representation for QM. Applying the Born rule there is a natural assignment of intensive values to all hermitian operators which are part of the decomposition of the quantum wave function, $\Psi$. This assignment is independent of the context of inquiry, allowing to derive explicitly an intensive non-contextuality theorem \cite[section 7]{deRondeMassri18a}. Quantum contextuality appears then in a new light, as making reference to the epistemic incompatibility of measurements instead to the ontic incompatibility of properties. In this way we drop the idea that the choice of the basis can modify the physical structure of the system under investigation \cite[section 8]{deRondeMassri18a}. Second, following the ongoing contemporary experimental research which points to the idea that quantum superpositions must be considered as real existents in Nature (e.g., \cite{Blatter00, NimmrichterHornberger13}), our intensive approach to quantum probability provides an account of superpositions which goes beyond the mere reference to measurement outcomes and mathematical formalisms \cite{deRonde18a, deRondeMassri18b}. Third, the objectivist intensive interpretation of the Born rule allows to place on equal footing both pure states and mixtures escaping the present difficulties in the definition of {\it quantum entanglement}, and consequently, the understanding of quantum information processing itself. This is a point of outmost importance in order to approach an objective definition of entanglement \cite{deRondeMassri18c}.  

All the just mentioned motivations for our intensive approach to probability have also interesting consequences for a possible development of both QC and QCL. From a philosophical perspective, the intensive approach might allow a realist understanding of the quantum computational processes in terms of interactions that are really taking place in the world. This, instead of the instrumentalist account which simply argues that ``it somehow works'' so you should better ``shut up and calculate!''. From a theoretical perspective, the implementation of an intensive account of quantum probability might allow a deep reconsideration of the way in which both {\it inputs} and {\it outputs} are considered and represented in QC. Since all terms of a qubit are considered as elements of physical reality (quantified in intensive terms) both inputs and outputs might necessarily require a multi-target approach to QC and QCL ---such as the one proposed in \cite{S1}. Another important aspect is that the empirical study of intensive values is necessarily linked to a statistical analysis in which single measurement outcomes play no essential role. Single outcomes simply don't provide enough information to account for intensive values ---even in extreme cases where probability is equal to 1. Intensive certainty, when empirically tested, always requieres a repeated series of measurements. Finally, it is interesting to point to the fact that the explicit consideration of different possible basis in QC and QCL, which can be applied through the implementation of qdits \cite{S2}, might open the door to analyze the contextual aspect of quantum computations \cite{Howard14} right from the start, not only taking into account what happens within the computational process, but also by considering the {\it output} and the {\it input} as explicitly contextual (or basis dependent). The just mentioned aspects imply not only a shift in some kernel aspects of the modeling of QC and QCL, it also implies a different way of analyzing the obtained data.

\section{Conclusions: reliability and probability}

In this paper we have discussed and analyzed different interpretations of probability in the context of QM. Firstly, we have considered the classical statistical interpretation of probability in terms of ignorance about an actual state of affairs. Secondly, the subjectivist  interpretation of probability which makes reference to the agents' rational choice when betting on a specific outcome. And thirdly, we provided an interpretation of objective probability which allows us to discuss the quantum formalism as making reference to an objective state of affairs described in intensive terms. As we have shown, while in orthodox QC and QCL, probability is used and understood as making reference to the possibility of observing different measurement outcomes, within the intensive approach to QC and QCL it is possible to understand quantum probability as making reference to an objectively described state of affairs. This new intensive approach to quantum probability opens new possibilities and advantages which we plan to investigate in future works. Here we point to the fact that such a proposal justifies the requirement of a multi-target statistical analysis.

\section*{Acknowledgements} 

We would like to thank to anonymous referees for their careful reading and insightful comments and criticism which have improved the article substantially. This work was partially supported by the following grants: FWO project G.0405.08 and FWO-research community W0.030.06. CONICET RES. 4541-12 and the Project PIO-CONICET-UNAJ (15520150100008CO) ``Quantum Superpositions in Quantum Information Processing'', Horizon 2020 program of the European Commission: SYSMICS project, number: 689176, MSCA-RISE-2015,  Fondazione Banco di Sardegna project ``Science and its Logics'' (CUP F72F16003220002"), RAS project ``Time-logical evolution of correlated microscopic systems" CRP 55, L.R. 7/2007 (2015) and Fondazione Banco di Sardegna project ``Strategies and Technologies for Scientific Education and Dissemination" (CUP F71I17000330002).


\begin{thebibliography}{1}

\bibitem{Accardi82} Accardi, L. (1982). Foundations of quantum
probability. {\it Rend. Sem. Mat. Univ. Politecnico di Torino},
249-270.

\bibitem{AertsSassoli17} Aerts D. $\And$ Sassoli de Bianchi, M. (2917). Do spins have directions? {\it Soft Computing}, {\it 21}.

\bibitem{AKN} Aharanov, D., Kitaev, A. $\And$ Nisan, N. (1997). Quantum circuits with mixed states'', {\it Proc. 13th Annual ACM Symp. on Theory of Computation}, STOC,
 20-30.
 
\bibitem{AuffevesGrangier15} Auffeves, A. $\And$ Grangier, P. (2015). A simple derivation of Born's rule with and without Gleason's theorem. Preprint. (arXiv:1505.01369) 

\bibitem{Blatter00} Blatter, G. (2000). Schrodinger's cat is now fat. {\it Nature}, {\bf 406}, 25-26.

\bibitem{Bohm53} Bohm, D. (1953). Proof That Probability Density Approaches $|\psi|^{2}$ in Causal Interpretation of the Quantum Theory. {\it Physical Review}, {\bf 89}, 458-466.

\bibitem{Bohr35} Bohr, N. (1935). Can Quantum Mechanical Description of Physical Reality be Considered Complete?'', {\it Physical Review}, {\bf  48}, 696-702.

\bibitem{Bub97} Bub, J. (1997). {\it Interpreting the Quantum World}, Cambridge University Press, Cambridge.

\bibitem{Bub17} Bub, J. (2017). Quantum Entanglement and Information. In {\it The Stanford Encyclopedia of Philosophy (Spring 2017 Edition)}, Edward N. Zalta (ed.).

\bibitem{DGG} Dalla Chiara, M.L., Giuntini, R. $\And$ Greechie, R. (2004). {\it Reasoning in Quantum Theory, Sharp and Unsharp Quantum Logics}, Kluwer, Dordrecht-Boston-London.

\bibitem{DGS} Dalla Chiara, M.L., Giuntini, R. $\And$ Sergioli, G. (2013). Probability in Quantum Computationa and in Quantum Computational Logic'', {\it Mathematical Structures in Computer Science},14, Cambridge University Press.

\bibitem{DGLNS} Dalla Chiara, M.L., Giuntini, R., Leporini, R., Negri, E. $\And$ Sergioli, G. (2015). Quantum Information, Cognition and Music", {\it Frontiers in Psychology}, 25, 1.

\bibitem{DGLS} Dalla Chiara, M.L., Giuntini, R., Leporini, R. $\And$ Sergioli, G. (2016). Holistic logical arguments in quantum computation", {\it Mathematica Slovaca}, 66, 2.

\bibitem{DawidThebault14} Dawid, R. $\And$ Thebault, P.Y. (2014). Against the empirical viability of the Deutsch-Wallace-Everett approach to quantum mechanics. {\it Studies in History and Philosophy of Modern Physics}, {\bf 47}, 55-61.

\bibitem{deRonde16a} de Ronde, C. (2016). Probabilistic Knowledge as Objective Knowledge in Quantum Mechanics: Potential Immanent Powers instead of Actual Properties'. In D. Aerts, C. de Ronde, H. Freytes and R. Giuntini (Eds.), {\it Probing the Meaning of Quantum Mechanics: Superpositions, Semantics, Dynamics and Identity}, pp. 141-178, World Scientific, Singapore.

\bibitem{deRonde17} de Ronde, C. (2017). Causality and the Modeling of the Measurement Process in Quantum Theory. {\it Disputatio}, {\bf 9}, 657-690.

\bibitem{deRonde18a} de Ronde, C. (2018). Quantum Superpositions and the Representation of Physical Reality Beyond Measurement Outcomes and Mathematical Structures. {\it Foundations of Science}, {\bf 23}, 621-648.

\bibitem{deRonde19} de Ronde, C. (2019). Immanent Powers versus Causal Powers (Propensities, Latencies and Dispositions) in Quantum Mechanics. In D. Aerts, M.L. Dalla Chiara, C. de Ronde and D. Krause (Eds.), {\it Probing the Meaning of Quantum Mechanics. Information, Contextuality, Relationalism and Entanglement}, pp. 121-157, World Scientific, Singapore.

\bibitem{deRondeMassri18a} de Ronde, C. $\And$ Massri, C. (2018). The Logos Categorical Approach to Quantum Mechanics: I. Kochen-Specker Contextuality and Global Intensive Valuations. {\it International Journal of Theoretical Physics}, DOI: 10.1007/s10773-018-3914-0.

\bibitem{deRondeMassri18b} de Ronde, C. $\And$ Massri, C. (2018). The Logos Categorical Approach to Quantum Mechanics: II. Quantum Superpositions and Measurement Outcomes. {\it International Journal of Theoretical Physics}, submitted (arXiv:1802.00415).

\bibitem{deRondeMassri18c} de Ronde, C. $\And$ Massri, C. (2018). The Logos Categorical Approach to Quantum Mechanics: III. Relational Potential Coding and Quantum Entanglement Beyond Collapses, Pure States and Particle Metaphysics. Preprint (arXiv:1807.08344).

\bibitem{Deutsch99} Deutsch, D. (1999). Quantum Theory of Probability and Decisions. {\it Proceedings of the Royal Society of London A455}, 3129-3137.

\bibitem{Deutsch04} Deutsch, D. (2004). {\it The Beginning of Infinity. Explanations that Transform the World}, Viking, Ontario. 

\bibitem{De} Devitt, S.J. (2016). Performing quantum computing experiments in the cloud.
{\it Physical Review A}, {\bf 94}, 032329.

\bibitem{Dieks07} Dieks, D. (2007). Probability in modal interpretations of quantum mechanics. {\it Studies in History and Philosophy of Science Part B}, {\bf 38}, 292-310.

\bibitem{DF} Domenech, G. $\And$ Freytes, H. (2006). Fuzzy propositional Logic associated with Quantum Computational Gate. {\it International Journal of Theoretical Physics}, {\bf 45}, 228-261.

\bibitem{EPR} Einstein, A., Podolsky, B. $\And$ Rosen, N. (1935). Can Quantum-Mechanical Description be Considered Complete? {\it Physical Review}, {\bf 47}, 777-780.

\bibitem{Einstein34} Einstein, A. (1934). The Herbert Spencer lecture, delivered at Oxford, June 10, 1933. Published in Mein Weltbild, Amsterdam: Querida Verlag.

\bibitem{Evans78} Evans, G. (1978). Can there be Vague Objects? {\it Analysis}, {\bf 38}, 208.

\bibitem{FD} Freytes, H. $\And$  Domenech, G. (2013). Quantum computational logic with mixed states. {\it Mathematical Logic Quarterly}, {\bf 59}, 27-50.

\bibitem {FSA} Freytes, H., Sergioli, G. $\And$ Aric\'o, A. (2010). Representing continuous t-norms in quantum computation with mixed states. {\it Journal of Physics A}, {\bf 43}, no.46.

\bibitem{FuchsPeres00} Fuchs, C.A. $\And$ Peres, A. (2000). Quantum theory needs no `interpretation'. {\it Physics Today}, {\bf 53}, 70.

\bibitem{QBism13} Fuchs, C.A., Mermin, N.D. $\And$ Schack, R. (2014). An introduction to QBism with an application to the locality of quantum mechanics. {\it American Journal of Physics}, {\bf 82}, 749.

\bibitem{GUD1} Gudder, S. (2003). Quantum computational logic. {\it International Journal of Theoretical Physics}, {\bf 42}, 39-47.

\bibitem{Healey} Healey, R. (2018). Quantum Theory and the Limits of Objectivity. {\it  Foundations of Physics}, {\bf 48}, 568-1589. 

\bibitem{Heis73} Heisenberg, W. (1973). Development of Concepts in
the History of Quantum Theory. In J. Mehra (Ed.), {\it The Physicist's Conception
of Nature}, pp. 264-275, Reidel, Dordrecht.

\bibitem{H} Hirvensalo, M. (2001). Quantum Computing. {\it Natural Computing Series}, Springer.

\bibitem{Howard14} Howard, M., Wallman, J., Veitch, V., $\And$ Emerson J. (2014). Contextuality supplies the magic for quantum computation. {\it Nature}, {\bf 510}, 351-355. 

\bibitem{Kalmbach} Kalmbach, G. (1986). Measures and Hilbert lattices. World Scientific, Singapore.

\bibitem{Kent14} Kent, A. (2014). Quantum Theory's Reality Problem. {\it Aeon}. (arXiv:1807.08410).

\bibitem{KSV} Kitaev, A.Y., Shen A. $\And$ Vyalyi, M.N. (2002). Classical and Quantum Computation. {\it Graduate Studies in Mathematics- AMS}, 47.

\bibitem{KS} Kochen, S. $\And$ Specker, E. (1967). On the problem of Hidden Variables in Quantum Mechanics. {\it Journal of Mathematics and Mechanics}, {\bf 17}, 59-87. 

\bibitem{Kolmogorov} Kolmogorov, A. (1950). {\it Grundbegriffe der Wahrscheinlichkeitrechnung, Ergebnisse Der Mathematik}; translated as: {\it Foundations of Probability}, Chelsea Publishing Company.

\bibitem{LS} Ledda, A. $\And$ Sergioli, G. (2010). Towards Quantum Computational Logics. {\it International Journal of Theoretical Physics}, {\bf 49}, 46.

\bibitem{Mermin98} Mermin, D. (1998). What is quantum mechanics trying to tell us? {\it 
American Journal of Physics}, {\bf 66}, 753-767.

\bibitem{Mermin14} Mermin, D. (2014). Why QBism is not the Copenhagen interpretation and what John Bell might have thought of it. Preprint (arXiv:1409.2454).

\bibitem{NC} Nielsen, M.A. $\And$ Chuang, I. (2000). Quantum Computation and Quantum Information. Cambridge University Press.

\bibitem{NimmrichterHornberger13} Nimmrichter, S. $\And$ Hornberger, K. (2013). Macroscopicity of Mechanical Quantum Superposition States. {\it Physical Review Letters}, {\bf 110}, 160403.

\bibitem{Pauli94} Pauli, W. (1994). {\it Writings on Physics and Philosophy} Enz, C. and von Meyenn, K. (Eds.), Springer-Verlag, Berlin.

\bibitem{Peres} Peres, P. (1995). {\it Quantum Theory: Concepts and Methods}, Fundamental Theories of Physics series, Klwer Academic Publisher, vol. 72, 192-209. 

\bibitem{Pitowsky94} Pitowsky, I. (1994). George Boole's `Conditions of Possible Experience' and the Quantum Puzzle. {\it The British Journal for the Philosophy of Science}, {\bf 45}, 95-125.

\bibitem{Primas99} Primas, H. (1999). Basic elements and problems
of probability theory. {\it Journal of Scientific Exploration}, {\bf 13}, 579-613.

\bibitem{Schr35} Schr\"odinger, E. (1935). The Present Situation in Quantum Mechanics. {\it Naturwiss}, {\bf 23}, 807. Translated to english in: Quantum Theory and Measurement,  Wheeler JA, Zurek WH (Eds.), 1983, Princeton University Press, Princeton.

\bibitem{SelleriTorizzi81} Selleri, F. $\And$ Tarozzi, G. (1981). Quantum Mechanics Reality and Separability'. {\it Rivista del Nuovo Cimento}, {\bf 4}, 1-53.

\bibitem{S1} Sergioli, G. (2019). Towards a Multi Target Quantum Computational Logic. {\it Foundations of Sciences}, DOI: 10.1007/s10699-018-9569-8.

\bibitem{S2} Sergioli, G. (2019). A matrix representation of quantum circuits over non-adjacent qudits. {\it International Journal of Theoretical Physics}, DOI: 10.1007/s10773-019-04051-5.

\bibitem{SGP} Sergioli, G., Giuntini, R. $\And$ Paoli, F. (2011). Irreversibility in Quantum Computational Logics. {\it Applied Mathematics and Information Sciences}, 5, 171-191.

\bibitem{Svozil17} Svozil, K. (2017). Classical versus quantum probabilities and correlations. Preprint (arXiv:1707.08915).

\bibitem{TA} Tarasov, V. (2002). Quantum computer with Mixed States and Four-Valued Logic. {\it Journal of Physics A}, {\bf 35}, 5207-5235.

\bibitem{Wallace07} Wallace, D. (2007). Quantum probability from subjective likelihood: Improving on Deutsch's proof of the probability rule. {\it Studies in the History and
Philosophy of Modern Physics}, {\bf 38}, 311-332.

\bibitem{WZ} Wheeler, J. $\And$ Zurek, W. Eds. (1983), {\it Theory and Measurement}. Princeton University Press, Princeton.
\end{thebibliography}
\end{document}